**Distinct Competing Ordered ν=2 States in Bilayer Graphene**

J. Velasco Jr.[1*§], Y. Lee[1*], Fan Zhang[2*], Kevin Myhro[1], David Tran[1], Michael Deo[1], Dmitry Smirnov[3], A. H. MacDonald[4], Chun Ning Lau[1†]

[1] Department of Physics and Astronomy, University of California, Riverside, CA 92521
[2] Department of Physics and Astronomy, University of Pennsylvania, Philadelphia, PA 19104
[3] National High Magnetic Field Laboratory, Tallahassee, FL 32310
[4] Department of Physics, University of Texas at Austin, Austin, TX 78712
[§] Current address: Department of Physics, University of California, Berkeley, CA 94709
[*] These authors contribute equally to this manuscript.

**Because of its large density-of-states and the $2\pi$ Berry phase near its low-energy band-contact points[1-6], neutral bilayer graphene (BLG) at zero magnetic field ($B$) is susceptible to chiral-symmetry breaking, leading to a variety of gapped spontaneous quantum Hall states distinguished by valley and spin-dependent quantized Hall conductivities.[7-34] Among these, the layer antiferromagnetic state,[35-37] which has quantum valley Hall (QVH) effects of opposite sign for opposite spins, appears to be the thermodynamic ground state. Though other gapped states have not been observed experimentally at $B$=0, they can be explored by exploiting their adiabatic connection to quantum Hall states with the same total Hall conductivity $\sigma_H$. In this paper, by using a magnetic field to select filling factor ν=2 states with $\sigma_H=2e^2/h$, we demonstrate the presence of a quantum anomalous Hall (QAH) state for the majority spin, and a Kekulé state with spontaneous valley coherence and a quantum valley Hall state for the minority spin in BLG. By providing the first spectroscopic mapping of spontaneous Hall states at ν=2, our results shed further light on the rich set of competing ordered states in BLG.**

In the $B=0$ gapped broken symmetry states of BLG, each spin-valley flavor spontaneously transfers charge between layers.[10,11] When spin is ignored only two classes of gaped states can be distinguished: (i) ones in which electrons in opposite valleys are polarized to opposite layers, producing a QAH state with broken time-reversal symmetry and orbital magnetization[7-9], and (ii) states in which electrons in opposite valleys are polarized to the same layer, yielding a QVH state with broken inversion symmetry and $\sigma_H=0$ [7-9]. For spinful electrons, symmetry-broken states further include the layer antiferromagnet (LAF) and the quantum spin Hall state with two-terminal conductivity 0 and $4e^2/h$, respectively. Electronic configurations of these states are summarized in Fig. 1a. So far only the LAF state, which has QVH states with opposite layer polarization signs for opposite spins[9,10,12,13,20,21,23,24,28,29], and (at finite $E_\perp$) the layer polarized state, which has QVH states with spin-independent layer polarization, have been observed experimentally.[35-38] Other spontaneous quantum Hall states, and competing states with Kekulé order, *i.e.* with spontaneous coherence between valleys, have so far eluded experimental observation. The experiments we describe are motivated by the sensitivity of competing order in BLG to both $E_\perp$ and $B$ fields.

Spontaneous Hall states can be explored by taking advantage of their adiabatic connection to QH states at integer filling factors within bilayer graphene's eightfold degenerate $N=0$ Landau levels (LLs). We use transport spectroscopy[35,36,39] to study high quality dual gated suspended BLG devices at $\nu=\pm2$ [36,38,40-45] as a function of $B$ and $E_\perp$. Our results reveal two distinct states: (1). Phase I is fully resolved only near $E_\perp=0$ and large $B$, and has a relatively small gap that extrapolates to 0 at $B=0$. This phase is expected[8] to be spontaneously spin-polarized and to have coherence between valleys (layers), *i.e.* with Kekulé valley order. (2). Phase II appears only at finite $E_\perp$, but can be stabilized at a much smaller $B$ and has a much larger gap with a finite $B=0$ intercept, suggesting that this state survives to anomalously weak $B$. Because finite B couples to the spin and orbital moments and finite $E_\perp$ favors layer (valley) polarization, this state is expected[8] to form a majority-spin QAH state but minority-spin QVH state. It follows that its charge, spin, and valley Hall conductivities are all quantized to $2e^2/h$. The anomalous stability of phase II at small $B$ is consistent with the stability of QVH (QAH) state at large $E_\perp$ (finite $B$). Our results represent the first spectroscopic mapping of the exotic behavior hosted in BLG at filling factor $\nu=2$. By shedding light on the competition between ordered states in BLG, we provide a roadmap for studying the spontaneous symmetry breaking in BLG and its thicker chiral cousins.[8,46,47]

---

† Email: jeanie.lau@ucr.edu

BLG sheets are exfoliated onto Si/SiO$_2$ substrates, and identified via color contrast in an optical microscope and through Raman spectroscopy[48]. Suspended dual-gated devices (Fig. 1b) with mobilities as high as 150,000 cm$^2$/Vs were measured in He$^3$ refrigerators. Here we present data from two different devices (device 1 and 2) with field effect mobilities 80,000 and 40,000 cm$^2$/Vs, respectively. All measurements were taken at temperature $T$=260mK.

Fig. 2a plots the 2-terminal differential conductance $G$ of device 1 in units of $e^2/h$ at $B=3.5$T as a function of charge density $n$ and out-of-plane electric field $E_\perp$. The QH plateaus at $\nu=0, \pm1, \pm2$ and $\pm4$ appear as blue, green, yellow, and brown color bands, respectively. Most interestingly, the resolution of $\nu=\pm2$ QH states depends on $E_\perp$. This can be seen in the $G(\nu)$ traces in Fig. 2b: near $E_\perp=0$, only the QH plateaus at $\nu=0$ and $\pm4$ are fully resolved; in contrast, at larger $E_\perp=-21$ mV/nm, the $\nu=\pm2$ plateaus are clearly visible. The resolution of the $\nu=\pm2$ states is abrupt: at $\nu=2$, $G(E_\perp)$ stays at $4e^2/h$ for small $E_\perp$, but drops sharply to a $2e^2/h$ plateau at a well-defined critical value $E_{\perp c}$. This critical $E_{\perp c}$ is ~ 10 mV/nm at $8$T, and is only weakly dependent on $B$, with a slope ~0.72 mV/nm/T (Fig. 2c).

The above observations reflect unprecedented sample quality, but agree with previous studies in which the $\nu=2$ state is fully resolved only in when $E_\perp$ is non-zero either in a controlled fashion in dual-gated devices[36,38] or inadvertently in singly-gated devices[35,37,42,45]. This suggests that the $\nu=2$ state observed at $B=3.5$T is layer polarized. The intriguing possibility of a $\nu=2$ state at $E_\perp=0$ in higher quality samples or in stronger fields has not been demonstrated previously. Fig. 2d displays $G(\nu)$ for device 2 at $E_\perp=0$ and two different $B$. Interestingly, a $\nu=2$ state is fully resolved only at $B=24$T, but no at lower field $B=10$T.

To further explore these two distinct $\nu=2$ QH states, we perform transport spectroscopy by using the source-drain bias $V$ as a spectroscopic tool[35,39]. The details of this measurement technique and its application to QH states can be found in ref. 36 and 39. Briefly, at each conductance plateau, BLG's Fermi level is pinned between the highest filled and the next unfilled LLs. Charges are carried by edge states, which are separated from the gapped bulk by a gap on the order of LL spacing. Increasing bias aligns the source electrode's Fermi level with the next unfilled LL and allows additional charge transport through the bulk. Consequently, near $V=0$, the device displays a conductance valley, whose half-width yields the LL gap $\Delta$. Fig. 3a-b display $G(V,\nu)$ at $B=3.5$T and $E_\perp=0$ and -14.4 mV/nm, respectively. At $E_\perp=-14.4$ mV/nm, the diamond at $v=-2$ is significantly larger than that at $E_\perp=0$, demonstrating an enlarged LL gap. Prominent zero-bias valleys appear in $G(V)$ traces (Fig. 3c-d, red lines), indicating that the $v=-2$ state (phase II) is fully resolved. In contrast, at $E_\perp=0$, the $G(V)$ traces display only small conductance dips superimposed on a peak at zero bias (Fig. 3c-d, blue traces), suggesting that phase I is only partially resolved.

The measured values of $\Delta_{v=-2}$ are shown in Fig. 3e. At $E_\perp=0$, the LL gap for phase I is $\Delta_I \sim 0.17$ meV/T with a zero intercept at $B=0$; in contrast, at $E_\perp=-14.4$ mV/nm, the gap for phase II is larger by more than a factor of 5, $\Delta_{II} \sim 1.0$ meV/T, with a non-vanishing $B=0$ intercept.

Thus, we observe two distinct QH states at $v=2$ experimentally. Phase I is only fully resolved at $E_\perp=0$ and large $B$, with a vanishing LL gap at $B=0$. The fact that a large $B$ is required to stabilize this phase has been anticipated in BLG QH ferromagnetism theory[5,8,22]. In the absence of interactions, the zero-energy LLs are 8-fold degenerate and QH effects occur only exhibit at $v=\pm 4$. Thus phase I arises from interactions. Because $N=0$ LL states in different valley are localized in different layers, there is an electrostatic energy cost of valley polarization. It is therefore energetically favorable to achieve a gap at $v=2$ by establishing coherence between valleys, i.e. Kekulé order. For this reason Kekulé order does appear at large $B$, even though it is absent at $B=0$. Our experiments demonstrate that the stability of the Kekulé state, as measured by its gaps, disappears as $B$ goes to zero.

In contrast to Phase I, Phase II is observed at anomalously small $B$ and large $E_\perp$, with a LL gap that extrapolates to a finite $B=0$ intercept. Its appearance at much smaller $B$ than Phase I is reminiscent of the spontaneous QH states at $B=0$[8,9]. This phase is adiabatically connected to the $B=0$ collinear ferromagnet ordered state with a majority spin QAH state and a minority spin QVH state[6,8,22]. Indeed phase II is only metastable[6,8] at $B=0$, most likely because it loses the ordering competition to the LAF state at $E_\perp=0$ and to the QVH state at $E_\perp \neq 0$. As observed here, however, phase II can be preferred in the presence of finite $B$ and $E_\perp$, since states with different total Hall conductivity are most stable at different carrier densities; moreover, this phase's energy is lowered by the orbital and spin coupling to $B$, and by the compensation of the Hartree energy cost of its layer polarization by $E_\perp$.

In terms of their layer characteristics, phase I has XY valley (layer) coherence order whereas phase II has layer-polarization Ising order. Schematic representations of these two phases are provided in Fig. 3f. The energy differences between these two phases arise mainly from (i) the Hartree energy of layer polarization and (ii) the difference between interlayer and intralayer exchange for the occupied $N=0$ LLs. Both differences scale as $(d/l_B)*(e^2/l_B) \sim B$, in agreement with the weak linear $B$ dependence of $E_{\perp c}$ in our BLG measurements, where the layer separation $d$ is much smaller than the magnetic length $l_B$.

We also note that data from singly-gated devices are similar to those at finite $E_\perp$: due to the presence of an inadvertently induced electric field $ne/2\varepsilon_0$ (here $\varepsilon_0$ is the permittivity of vacuum), the $v=\pm 2$ states in singly-gated devices are almost always in phase II and stabilized by $E_\perp$. Thus, Phase I has not been observed before and Phase II was observed accidentally.

Lastly, using transport spectroscopy, we explore the dependence of the LL gap on $E_\perp$. Fig. 4a plots $G(V, E_\perp)$ at $v=2$. The most striking feature is the red region at the center of the plot, i.e. at small $E_\perp$, where $G \sim 4e^2/h$, surrounded by blue-white regions where $G \sim 2e^2/h$. The abrupt transition between the 2 regions (Fig. 4b) is the same as that observed in Fig. 2c. At $E_\perp=0$, $G(V)$ displays a narrow conductance dip at $V=0$, with half-width $\sim 1.6$ mV, corresponding to $\Delta_I$ (Fig. 4c). At large $E_\perp=-35$ mV/nm, the $G(V)$ trace is significantly different, with much wider conductance valley and half-width of *15* meV. We thus take the half-width of wider valley to be $\Delta_{II}$. The dependence of $\Delta_{II}$ on $E_\perp$ is shown in Fig. 4d. It increases with $E_\perp$ of both polarity, though with slight asymmetry. Interestingly, the wider conductance valley, which appears as the white curves superimposed on top of the blue background, extend into $|E_\perp| < E_{\perp c}$, i.e. the narrow dip co-exists with wider valley, suggesting co-existence of both states near $E_{\perp c}$.

In conclusion, we utilized high quality dual gated suspended BLG devices to explore the spontaneous symmetry breaking physics near charge neutrality. We resolved two distinct *v=2* QH states: phase I is fully resolved only at $E_\perp=0$ and large *B*, and is likely a Kekulé state with inter-layer and inter-valley coherence; phase II is observed at small *B* and large $E_\perp$. Our measurements demonstrate that it is metastable at *B=0*. Our data represent the first spectroscopic mapping of the exotic competing orders in BLG, and pave the way for studies of other symmetry-broken states at different filling factors in BLG or in the much less explored TLG[8,49]. Our study motivates future in-plane *B* measurements[9,50,51] to realize more complete control of spin, layer, and orbital degrees of freedom.

Acknowledgement
We thank Philip Kratz for assisting sample preparation, and O. Vafek for helpful discussion. This work is supported in part in part by DMEA H94003-10-2-1003, NSF DMR/1106358, ONR and the FAME center that is one of the six STARNET centers and an SRC program sponsored by MARCO and DARPA. YL acknowledges the support by DOE. Part of this work was performed at NHMFL that is supported by NSF/DMR-0654118, the State of Florida, and DOE.

**Fig. 1. (a).** Schematic diagram of electronic configurations of the states stabilized by electric field and magnetic field. **(b).** SEM image of a dual-gated BLG device.

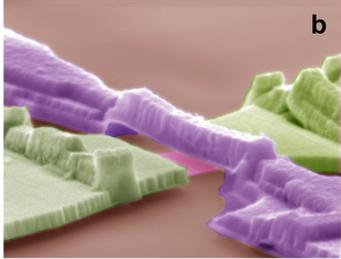

| a | K↑ | K↓ | K'↑ | K'↓ | G ($e^2/h$) | Layer polarization |
|---|---|---|---|---|---|---|
| Quantum Anomalous Hall (QAH) | T | T | B | B | 4 | No |
| Layer Antiferromagnet | T | B | T | B | 0 | No |
| Quantum Spin Hall | T | B | B | T | 4 | No |
| Quantum Valley Hall | T | T | T | T | 0 | Yes |
| ν=2 QH state | T | T | T | B | 2 | Yes |

**Fig. 2.** Data from Device 1 and 2. **(a).** $G(E_\perp, n)$ of Device 1 at $B$=3.5T. **(b).** Line traces $G(\nu)$ at $B$=3.5T and $E_\perp$=0 (red dotted line) and $E_\perp$=-21 mV/nm (blue solid line), respectively. **(c).** Line trace $G(E_\perp)$ at $B$=3.5T and filling factor $\nu$=2 (density $n$=1.7x10$^{15}$ cm$^{-2}$). **Inset:** Critical electric field $E_c$ vs. $B$ and linear fit to the data points with slope ~0.72 mV/nm/T. **(d).** $G(\nu)$ of device 2 at $E_\perp$=0 and $B$=10T and 24T, respectively.

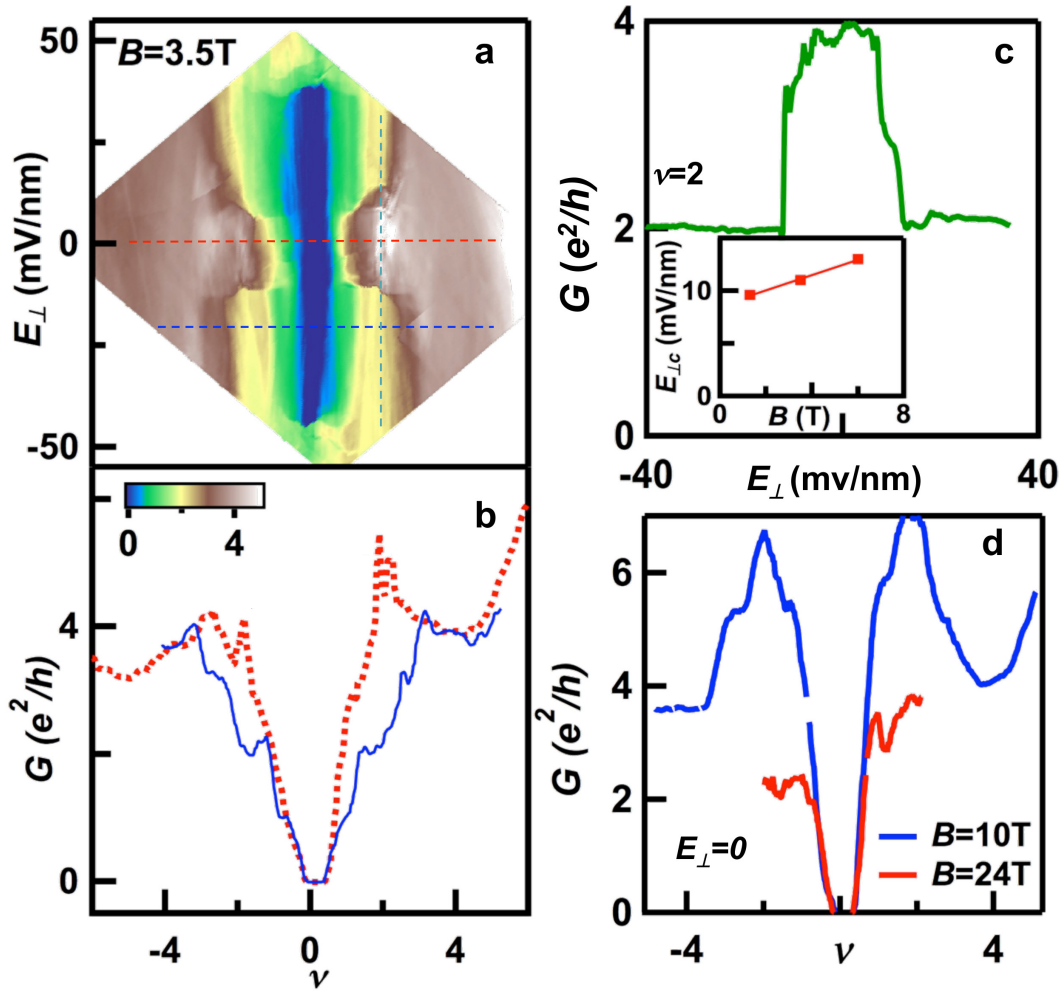

**Fig. 3. (a-b).** $G(V,n)$ at $E_\perp=0$ and -14.4 mV/nm, respectively. Both data are taken at $B$=3.5T. **(c).** Line trace $G(V)$ at $\nu$=-2 and $B$=3.5T. Blue and red traces are taken at $E_\perp=0$ and -14.4 mV/nm, respectively. **(d).** Similar data taken at $B$=6T. **(e).** Measured LL gap $\Delta(B)$ at $\nu$=2 and $E_\perp=0$ and -14.4 mV/nm, respectively. **(f).** Schematics of transitions between Phase I and Phase II at $\nu$=2. (T: top layer; B: bottom layer; S:symmetric state T+B; AS: anti-symmetric state T-B)

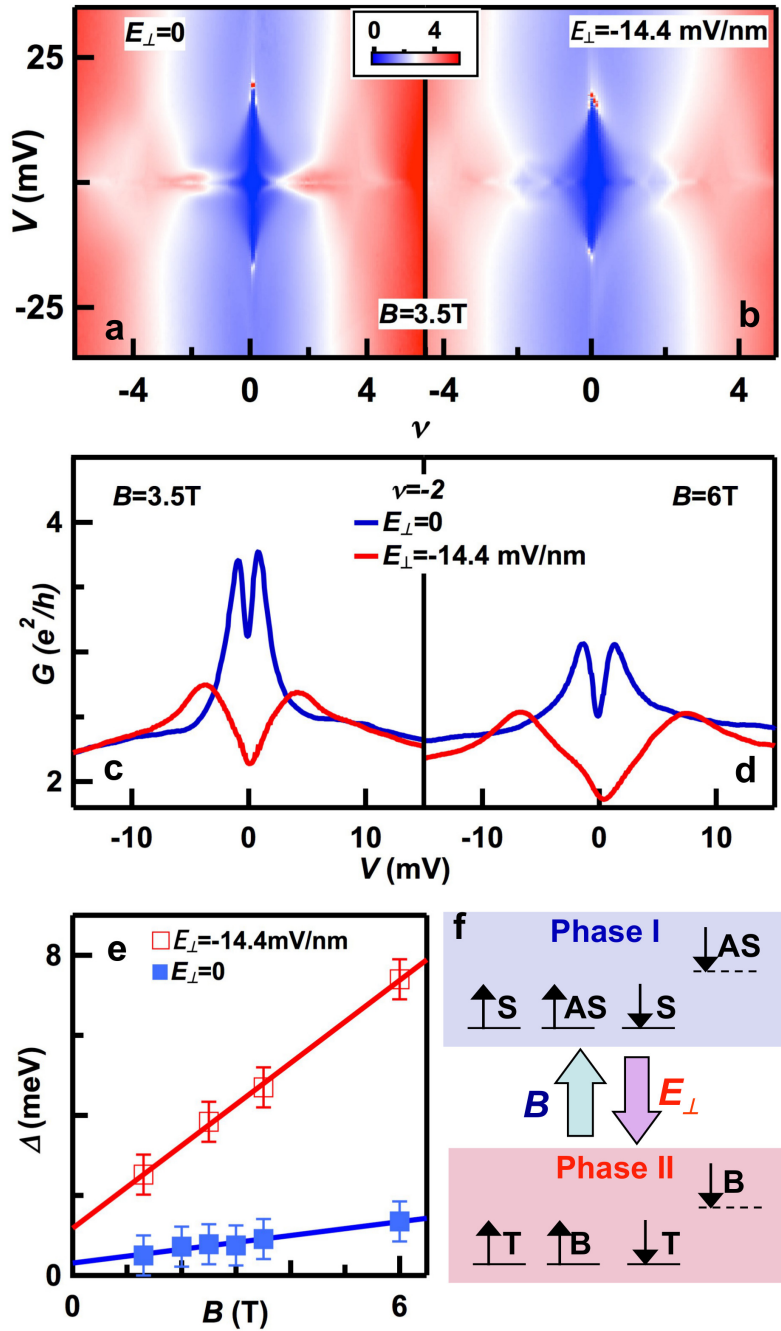

**Fig. 4.** Spectroscopy data from Device 2 at $B=10T$ and $v=2$. **(a).** $G(V, E_\perp)$ data. **(b).** Line traces $G(V)$ at $E_\perp=0$ (blue), -35 mV/nm (red) and 15 mV/nm (green). **(c).** $G(E_\perp)$ at $V=0$. **(d).** Gap of layer polarized $v=2$ state vs. $E_\perp$.

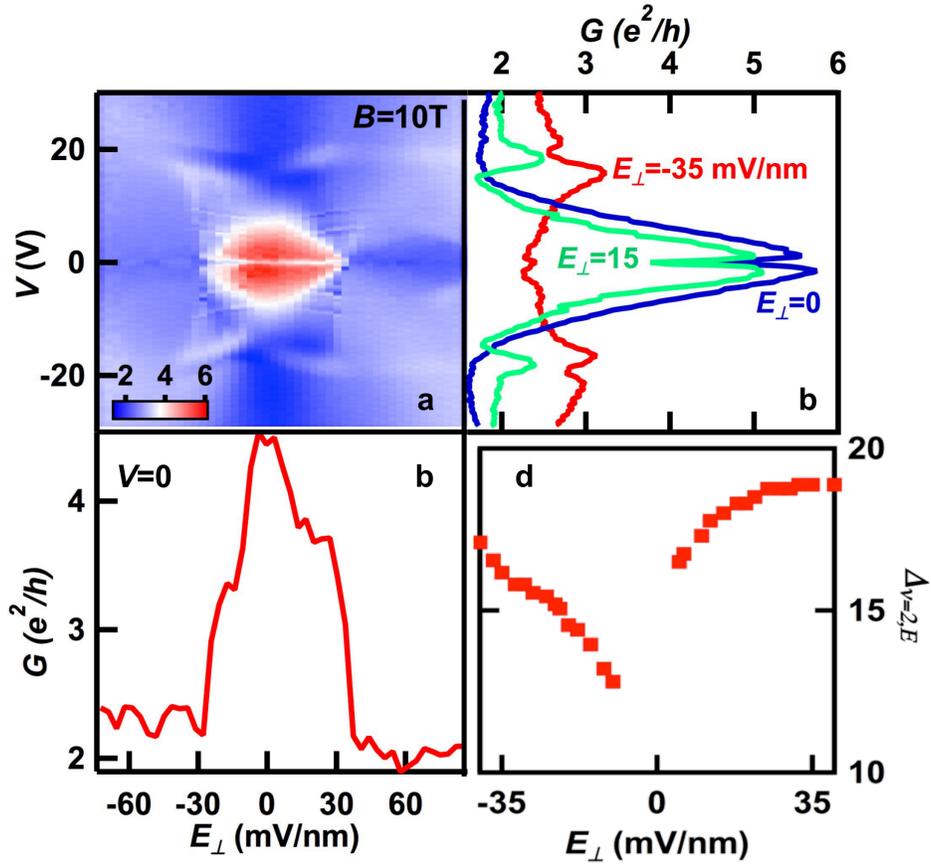